\documentclass[sigconf]{acmart}
\AtBeginDocument{%
  }

\setcopyright{none}
\copyrightyear{2026}
\acmYear{2026}
\acmDOI{XXXXXXX.XXXXXXX}

\usepackage{siunitx}
\begin{document}

\title{Operational Strategies for Non-Disruptive Scheduling Transitions in Production HPC Systems}


\author{Glen MacLachlan}
\authornote{Both authors contributed equally to this research.}
\email{maclach@gwu.edu}
\orcid{0000-0002-4805-6789}
\affiliation{%
  \institution{The George Washington University}
  \city{Washington}
  \state{DC}
  \country{USA}
}

\author{Joseph Creech}
\authornotemark[1]
\email{jcreech@gwu.edu}
\affiliation{%
  \institution{The George Washington University}
  \city{Washington}
  \state{DC}
  \country{USA}
}

\author{Rubeel Muhammad Iqbal}
\email{rubeel@gwu.edu}
\orcid{0009-0009-3572-0917}
\affiliation{%
  \institution{The George Washington University}
  \city{Washington}
  \state{DC}
  \country{USA}
}

\author{Clark Gaylord}
\email{cgaylord@gwu.edu}
\orcid{0000-0001-7459-0179}
\affiliation{%
  \institution{The George Washington University}
  \city{Washington}
  \state{DC}
  \country{USA}
}

\author{Jake Messick}
\email{jake_messick@gwu.edu}
\affiliation{%
  \institution{The George Washington University}
  \city{Washington}
  \state{DC}
  \country{USA}
}

\renewcommand{\shortauthors}{MacLachlan et al.}


\begin{abstract}
Migrating heterogeneous high-performance computing (HPC) systems to resource-aware scheduling introduces both technical and behavioral challenges, particularly in production environments with established user workflows. This paper presents a case study of transitioning a production academic HPC cluster from node-exclusive to consumable resource scheduling mid-lifecycle, without disrupting active workloads. We describe an operational strategy combining a time-bounded compatibility layer, observability-driven feedback, and targeted user engagement to guide adoption of explicit resource declaration. This approach protected active research workflows throughout the transition, avoiding the disruption that a direct cut-over would have imposed on the user community. Following deployment, median queue wait times fell from 277 minutes to under 3 minutes for CPU workloads and from 81 minutes to 3.4 minutes for GPU workloads. Users who adopted TRES-based submission exhibited strong long-term retention. These results demonstrate that successful scheduling transitions depend not only on system configuration, but on aligning observability, user engagement, and operational design.
\end{abstract}



\maketitle

\section{Introduction}

Efficient resource allocation in heterogeneous high performance computing (HPC) systems relies on scheduling policies that explicitly expose compute resources, including CPUs, memory, and GPUs, to both the scheduler and users, a challenge widely studied in parallel job scheduling~\cite{feitelson1997}.

This paper presents a production case study of migrating a heterogeneous HPC cluster from a node-exclusive to consumable
resource scheduling using Slurm Trackable Resources (TRES)~\cite{slurm_tres,yoo2003slurm}. The emphasis is on both the scheduler configuration and an operational strategy for deploying potentially disruptive changes in a live production environment while maintaining uninterrupted service.

The George Washington University's flagship HPC system, Pegasus~\cite{gwu_pegasus}, is a production cluster supporting a large interdisciplinary research community. At the time of migration, Pegasus comprised approximately 205 compute nodes (42 GPU nodes); 8,600 CPU cores; 280 TB aggregate system memory; 2 PB of storage; and 800 active researchers.

Over time, the node-exclusive scheduling model produced several structural inefficiencies including suboptimal node packing, increased queue wait times, and misaligned resource allocation patterns. This exposed both technical and behavioral challenges: submission patterns created measurable inefficiency, and transitioning researchers toward explicit resource declaration required deliberate operational design.

We present a production deployment strategy that combines phased transition with observability-driven feedback to enable this transition without disrupting active workloads. Aligning scheduler configuration with researcher-facing operational practices leads to substantial improvements in queue wait times, resource utilization, and sustained adoption of explicit resource requests. To address the technical component, the cluster migrated to Slurm’s Trackable Resources model with cgroup enforcement and GPU accounting through Slurm’s Generic Resource (GRES) framework~\cite{yoo2003slurm, slurm_gres}. While the technical implementation itself is well documented, the operational challenge centered on user adoption. The remainder of the paper focuses on the transition strategy, observability mechanisms, and operational lessons learned during the migration.

\section{Legacy Model Friction and Transition Strategy}

Under Pegasus's original node-exclusive scheduling discipline, the only mechanism available to differentiate resource classes was partition assignment. Over time, this produced a proliferation of specialized partitions: high-memory nodes, GPU variants, debug queues, visualization nodes, and high-throughput configurations. Each additional partition added incrementally to expose resource capabilities, while also increasing the perceived complexity of the submission environment for users. Under consumable resource scheduling, such distinctions are expressed through per-job resource requests rather than partition identity. The resulting complexity reinforced several behaviors:

\begin{itemize}
    \item Partition selection based on inferred resource availability rather than resource-as-declaration
    \item GPU allocation without corresponding GPU utilization
    \item Low core utilization per node due to implicit node exclusivity assumptions
\end{itemize}

Pre-migration telemetry revealed consistent patterns of low CPU packing efficiency, frequent allocation of GPU resources without utilization, and queue congestion driven by node exclusivity. These inefficiencies are consistent with known limitations of node-exclusive scheduling, including resource fragmentation and suboptimal utilization. Similar inefficiencies have been observed in large-scale GPU cluster environments, where mismatches between requested and actual resource usage lead to under-utilization~\cite{jeon2019analysis}. These observations motivated the need for a transition strategy that addressed both scheduling policy and user submission behavior.

Slurm’s Trackable Resources (TRES) model, implemented via the \texttt{select/cons\_tres} plugin, is enabled or disabled cluster-wide~\cite{slurm_tres}. As a result, changes to the scheduling model must be applied uniformly, requiring operational strategies to support user transition in a live production environment. A direct cutover was therefore rejected in favor of an approach that would allow the research community to adapt without interruption to active workloads.

To reduce user friction, we implemented a 90-day transition period during which legacy submission patterns were permitted but monitored. This approach preserved continuity of service while allowing users to adapt to explicit resource declarations. Lightweight wrapper scripts translated legacy job submissions into resource-aware equivalents where possible, enabling existing workflows to continue functioning while guiding users toward explicit CPU, memory, and GPU requests. These wrappers were time-bounded by the 90-day window, giving users a structured path toward native TRES-based submission patterns.

During the transition, legacy submission patterns were permitted but monitored and users received targeted feedback on historical inefficiencies. To ease user adoption, documentation was produced that emphasized resource declaration over partition identity, and office hours and town halls contextualized the change in terms of throughput and fairness.

\section{Observability as Stabilization Mechanism}

System observability played a critical role in evaluating the impact of the scheduling migration. Monitoring combined Slurm~\cite{slurm} accounting data with Zabbix infrastructure telemetry~\cite{zabbix}. Slurm accounting records provided job-level telemetry including requested resources, allocated resources, job duration, and queue wait time. Zabbix agents collected node-level metrics including memory usage, GPU utilization, and CPU core utilization.

Operational metrics tracked included:

\begin{itemize}
\item Queue wait time percentiles by partition
\item CPU packing efficiency
\item GPU allocation alignment
\item Memory utilization relative to requested memory
\end{itemize}

\begin{figure}[t]
\centering
\includegraphics[width=\linewidth]{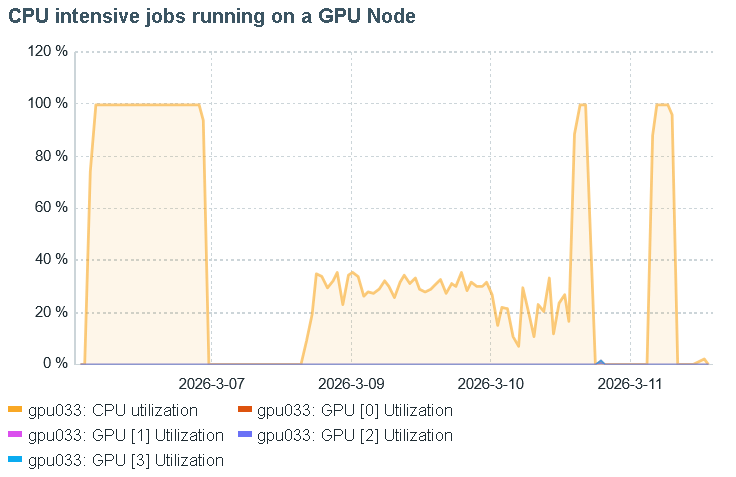}
\caption{CPU workloads running on a GPU node, under-utilizing the available GPU devices.}
\Description{Three separate CPU-intensive workloads running on a GPU node, under-utilizing the available GPU devices.}
\label{fig:cpu_gpu_node}
\end{figure}

These metrics enabled targeted feedback to users and supported rapid behavioral adjustment. For example, real-time Zabbix visualizations of GPU under-utilization were used during town halls and one-on-one consultations to demonstrate mismatches between requested and actual resource usage. This visibility helped researchers understand their own resource usage patterns and guided them toward more effective job submissions.

Figure~\ref{fig:cpu_gpu_node} shows three separate legacy jobs running on a single GPU node over a five-day period. Despite occupying a GPU-capable node, these jobs exhibit negligible or no GPU utilization, indicating that GPU resources were allocated but not consumed. This pattern reflects submission strategies that prioritized rapid job start times over appropriate resource matching, resulting in inefficient use of scarce GPU capacity. By contrast, under TRES-based scheduling, such mismatches are explicitly surfaced and discouraged, as resource requests must align with actual usage.

\section{Measured Effects}

Outcomes were evaluated across two phases: a pre-migration baseline and a 90-day transition window. During these phases, we tracked changes in queue wait times as well as user adoption and retention of TRES-based submission. Together, these metrics capture both system-level performance and the behavioral response to the scheduling transition.

\begin{figure}[t]
\centering
\includegraphics[width=\linewidth]{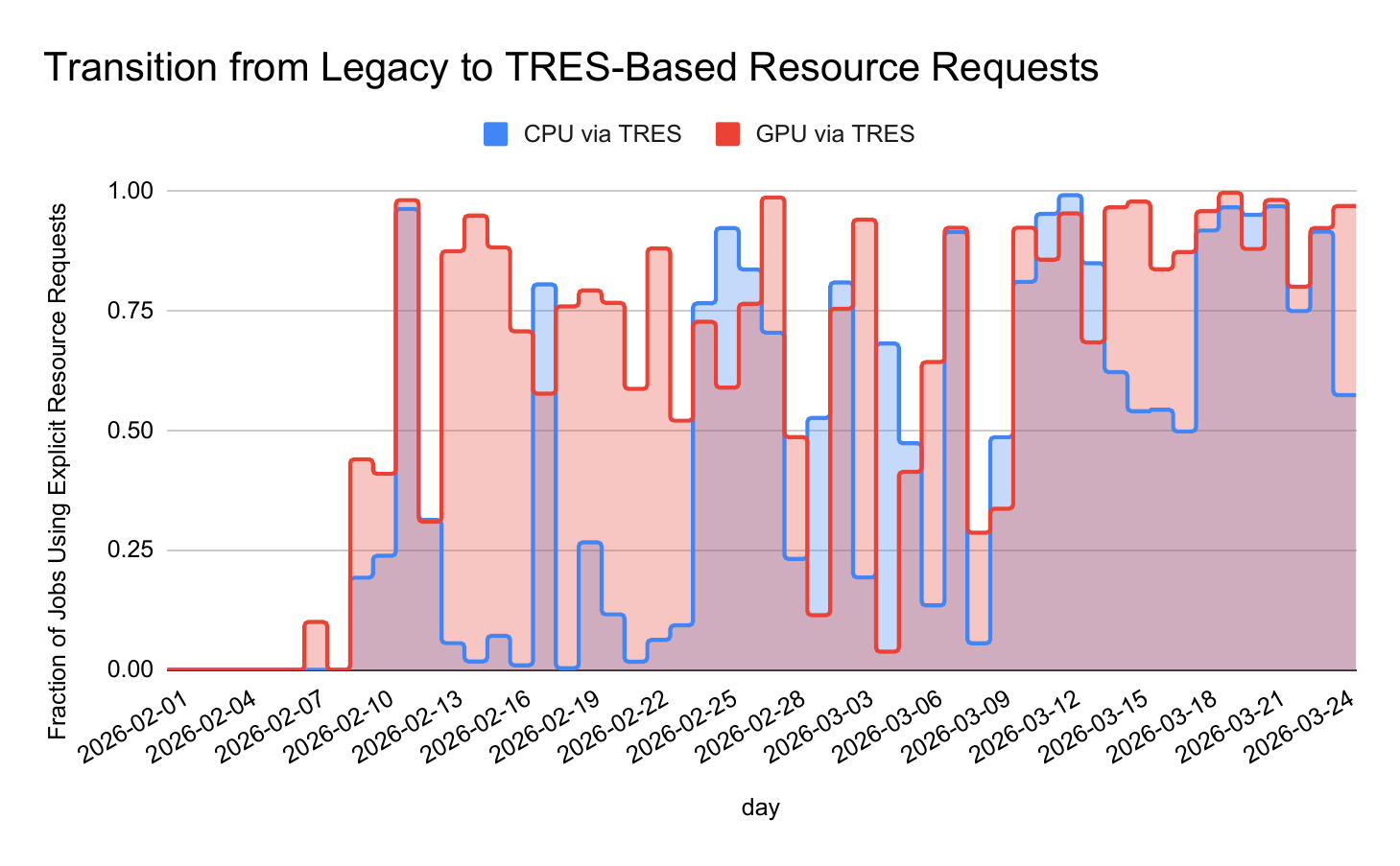}
\caption{Daily fraction of jobs using explicit resource declarations for CPU and GPU resources during the transition period. The data show progressive but non-monotonic adoption of TRES-based submission patterns, reflecting variability in workload composition and user behavior over time.}
\Description{A time series showing decline in legacy partition-based job submissions and increase in TRES-based submissions across a 90-day transition period.}
\label{fig:timeline}
\end{figure}

\subsection{Queue Wait Time}

\begin{table}[t]
\caption{Queue wait time comparison before and after TRES scheduling deployment. All wait times are reported in minutes.}
\label{tab:wait}
\centering
\small
\begin{tabular}{l
                S[table-format=6.0]
                S[table-format=4.2]
                S[table-format=4.2]
                S[table-format=4.2]}
\toprule
Partition & {Jobs} & {P50} & {P75} & {P90} \\
\midrule
Legacy CPU (Pre)  & 393188 & 277.18 & 800.69 & 3026.26 \\
\textbf{CPU (TRES)} & \bfseries 31563 & \bfseries 2.97 & \bfseries 106.12 & \bfseries 797.65 \\
Legacy CPU (Post) & 49428  & 294.80 & 513.19 & 642.53 \\
\midrule
Legacy GPU (Pre)  & 61077  & 81.05  & 1031.55 & 5567.88 \\
\textbf{GPU (TRES)} & \bfseries 14796 & \bfseries 3.40 & \bfseries 137.83 & \bfseries 461.60 \\
Legacy GPU (Post) & 6484   & 344.05 & 590.57 & 748.14 \\
\bottomrule
\end{tabular}
\end{table}

Table~\ref{tab:wait} shows queue wait time distributions across the transition. Job counts vary across categories due to differences in workload across each phase. For CPU workloads, TRES reduced median wait time from 277 minutes to under 3 minutes, with 90th-percentile latency decreasing from over 3,000 to under 800 minutes. GPU workloads improved similarly: median wait fell from 81 minutes to 3.4 minutes, and the 90th percentile from over 5,500 to under 500 minutes. Residual legacy GPU jobs submitted post-transition experienced substantially higher median wait times (344 minutes), indicating that continued reliance on legacy job submission is associated with substantially higher wait times as the user-base shifts toward TRES-based scheduling.

\subsection{Post-Adoption Behavior and Retention}

\begin{figure}[t]
\centering
\includegraphics[width=0.9\linewidth]{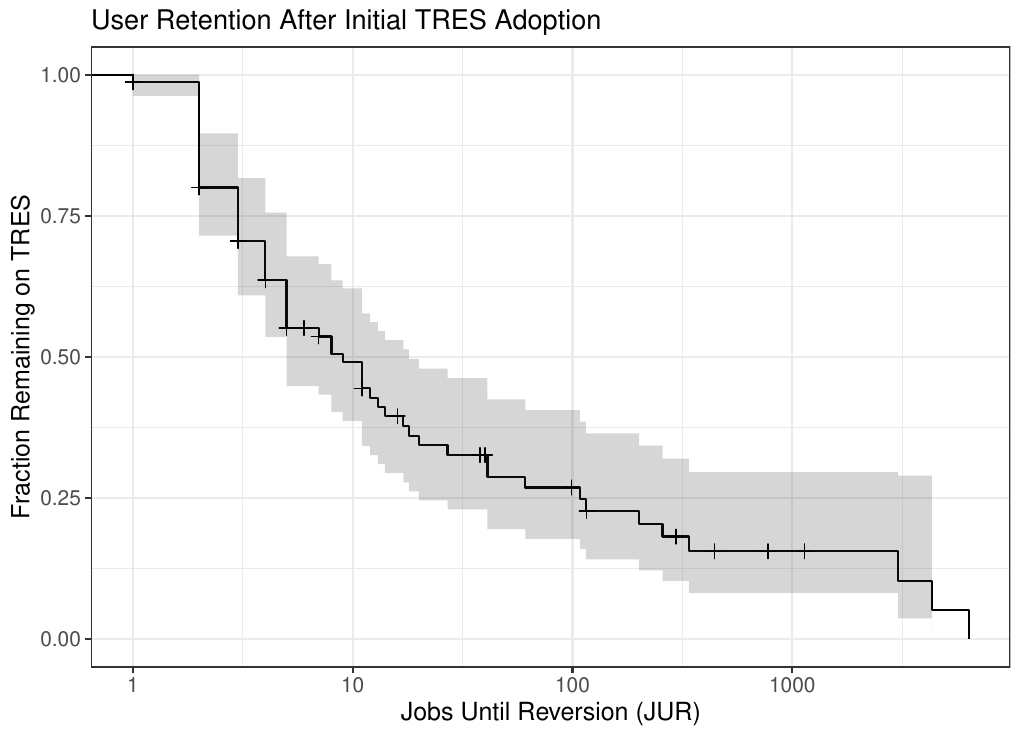}
\caption{Kaplan--Meier estimate of continued TRES usage as a function of Jobs Until Reversion to legacy (JUR) (log-scaled horizontal axis). The curve shows a sharp early decline followed by a plateau, indicating stable retention among users who persist beyond the initial jobs.}
\Description{Kaplan–Meier estimate of continued TRES usage as a function of number of jobs submitted after initial adoption, showing an early drop followed by a plateau.}
\label{fig:km}
\end{figure}

Figure~\ref{fig:timeline} shows the progression of TRES adoption over the transition period, highlighting both overall growth in explicit resource declarations and short-term variability. Figure~\ref{fig:km} shows a Kaplan–Meier~\cite{kaplan1958} estimate of continued TRES usage. Jobs until reversion (JUR) is defined as the number of jobs submitted after initial TRES adoption until first reversion to legacy, or until the end of the observation window if no reversion occurs. The curve shows a sharp early drop, with most reversion occurring within the first few jobs, followed by a gradual decline with wide confidence intervals, reflecting stable retention among the smaller risk set of high-volume users. This pattern is consistent with early submissions acting as a filtering stage: users who observe improved scheduling outcomes, including reduced median queue wait times (e.g., 3 minutes versus 277 minutes), tend to continue using TRES.

Approximately one week after deployment, a computational biologist provided unsolicited feedback: \textit{"I just wanted to pass along thanks for the implementation of the new scheduling system. It seems to have been a roaring success in terms of node utilization."} 

\section{Conclusion}
    
This deployment shows that introducing consumable resource scheduling in a production HPC environment is fundamentally an operational and behavioral challenge, not just a technical one; technical change alone exposes inefficiencies but does not resolve them.

Observability was a driver for change, allowing users to observe mismatches between requested and utilized resources, while the 90-day compatibility window gave researchers the time needed to adapt their workflows with minimal disruption.

These results indicate that successful scheduling transitions are best achieved through the coordinated use of telemetry, user engagement, and phased roll-out strategies.

\begin{acks}
The authors thank Fong Banh, Kai Leung Wong, and Kevin Weiss for their support in assisting users during the transition to
resource-aware scheduling, including guiding users through updated
submission practices and escalating operational issues.

\end{acks}
\bibliographystyle{ACM-Reference-Format}
\bibliography{references}

\end{document}